\documentstyle[aps,prb,floats,epsf]{revtex}

\begin{document}

\draft


\twocolumn[\hsize\textwidth\columnwidth\hsize\csname@twocolumnfalse%
\endcsname

\title{Impurity-induced spin polarization and NMR line broadening in 
underdoped cuprates}

\author{R.\ Kilian, S.\ Krivenko,\cite{KAZ} G.\ Khaliullin,\cite{KAZ} 
and P.\ Fulde}
\address{Max-Planck-Institut f\"ur Physik komplexer Systeme,
N\"othnitzer Strasse 38, D-01187 Dresden, Germany}

\date{Version date: \today}

\maketitle


\begin{abstract}
We present a theory of magnetic ($S=1$) Ni and nonmagnetic Zn impurities 
in underdoped cuprates. Both types of impurities are shown to induce 
$S=\case{1}{2}$ moments on Cu sites in the proximity of the impurity,
a process which is intimately related to the spin gap phenomenon in 
cuprates. Below a characteristic Kondo temperature, the Ni spin is
partially screened by the Cu moments, resulting in an effective impurity
spin $S=\case{1}{2}$. We further analyze the 
Ruderman-Kittel-Kasiya-Yosida-type response of 
planar Cu spins to a polarization of the effective impurity moments and 
derive expressions for the corresponding $^{17}$O NMR line broadening.
The peculiar aspects of recent experimental NMR data can be traced back
to different spatial characteristics of Ni and Zn moments as well 
as to an inherent temperature dependence of local antiferromagnetic
correlations.
\end{abstract}
\pacs{PACS numbers: 75.20.Hr, 74.25.Ha, 74.72.-h, 76.60.-k}]


\section{Introduction}
\label{SEC:INT}

The normal state of underdoped cuprates exhibits unusual magnetic properties
which are believed to be intimately related to the mechanism of
high-$T_c$ superconductivity. Most peculiar in this respect are the 
simultaneous occurrence 
of a magnetic pseudogap and the persistence of antiferromagnetic (AF) 
correlations as holes are doped into the antiferromagnetic insulator and 
the system becomes metallic. It is one of the most challenging theoretical
problems in the physics of high-$T_c$ cuprates to reconcile the gaplike 
features reminiscent of a spin liquid with the presence of antiferromagnetic 
correlations signaling the closeness of the system to a spin-ordered N\'eel 
state. Experimentally, insight into the nature of these anomalous
features can be gained by introducing impurities into the magnetically 
active Cu sites. A subsequent NMR probe on nuclei coupled to the CuO$_2$ 
planes yields information on the local magnetic structure. In this
paper we present a microscopic theory of the impurity-induced
local spin polarization of CuO$_2$ planes and its impact on the NMR linewidth.

Introducing magnetically active or inert impurities into underdoped 
cuprates leads, in both cases, to the formation of local magnetic moments.
Specifically, Cu ($d^9$) with an effective in-plane spin $S=\case{1}{2}$ 
can be replaced by Ni ($d^{8}$) with $S=1$ or Zn ($d^{10}$) with $S=0$.
Superconducting quantum interference device (SQUID) measurements of the 
macroscopic susceptibility \cite{MEN94} reveal 
an almost perfect $1/T$ Curie behavior. Recently, Bobroff {\it et al.} 
\cite{BOB97} presented NMR measurements on $^{17}$O for the underdoped 
compound $\text{YBa}_2(\text{Cu}_{1-x}M_x)_3\text{O}_{6.6}$, with $M$ = 
Zn or Ni. The polarization of Cu spins in the presence of impurities leads 
to a broadening of the NMR line. In contrast to the aforementioned SQUID
measurement, the linewidth displays a marked non-Curie behavior, 
indicating an inherent temperature dependence of the polarizability of 
CuO$_2$ planes. This was suggested by Morr {\it et al.} \cite{MOR98}
to be a clear indication for a temperature dependence of the AF
correlation length. Still another interesting 
observation can be made by comparing the two experiments: While the NMR 
study shows nonmagnetic Zn to have a more pronounced effect on the 
linewidth than Ni, measurements of the macroscopic susceptibility reveal 
a reversed effect.
Since only the NMR experiment is sensitive to a spatial variation 
of the spin polarization, a very different shape of the spin density induced 
by the two types of impurities can be inferred.

In the following, we present a microscopic theory of moments induced
by magnetic and nonmagnetic impurities in the spin gap phase of
underdoped cuprates.
We analyze the different nature of coupling between Cu and impurity
spins and derive expressions for the 
local spin polarization of CuO$_2$ planes. The presence of the spin gap
and of short-range AF correlations is shown to strongly modify the 
conventional Ruderman-Kittel-Kasiya-Yosida
(RKKY) picture. Finally, we derive expressions for the NMR
line broadening which account well for the peculiarities of the 
experimental data.

\section{Impurity Model}
\label{SEC:IMP}

The relevant physics of the CuO$_2$ planes of high-$T_c$ cuprates
is believed to be described by the large-$U$ Hubbard or $t$-$J$ model. 
The dualism between
itinerant charge motion and local electron interaction that is inherent to
these models can, in an approximate way, be captured by introducing
separate quasiparticles for spin and charge degrees of freedom.
Within this picture, the normal state of underdoped cuprates is viewed as
a phase in which spins form singlet pairs while coherence 
between holes that would eventually lead to superconductivity
has not been established. We follow this line of thinking but 
restrict ourselves to the magnetic sector of the Hilbert space. Our 
starting point is the spin-$\case{1}{2}$ AF Heisenberg model
$H_J = J\sum_{\langle ij \rangle} \bbox{s}_i \bbox{s}_j$.
Keeping in mind the presence of itinerant holes which prevent the
system from developing long-range magnetic order,
we treat this Hamiltonian within resonance valence bond (RVB) 
mean-field theory \cite{AND87}~-- this accounts well for the 
spin-liquid features of cuprates. The mean-field Hamiltonian is
\begin{equation}
H_{\text{RVB}} = -\sum_{\langle ij \rangle \sigma} \left( \Delta_{ij} 
f^{\dagger}_{i \sigma} f_{j \sigma}+ \text{H.c.} \right).
\label{HRVB}
\end{equation}
Original spin operators $\bbox{s}_i$ have been expressed in terms of 
fermionic operators by $\bbox{s}_i = \case{1}{2} \sum_{\sigma\sigma'}
\bbox{\tau}_{\sigma\sigma'} f^{\dagger}_{i\sigma} f_{i\sigma'}$ with 
Pauli matrix vector $\bbox{\tau} = (\tau^x,\tau^y,\tau^z)$.  
The local constraint prohibiting a double occupancy of sites has
been relaxed to a global one.
The mean-field bond parameter is $\Delta_{ij}=\Delta_{\bbox{\delta}}= J 
\sum_{\sigma}\langle f^{\dagger}_{i+\delta,\sigma} f_{i\sigma}\rangle^0$,
where $\langle\cdots\rangle^0$ is the expectation value that corresponds 
to Hamiltonian (\ref{HRVB}). The phase of this mean-field parameter
is yet undetermined and has to be chosen such as to 
resemble the experimental situation most closely. An appropriate
choice for the spin gap regime is the flux phase\cite{AFF88}
$\Delta_{\pm x} = i\Delta_{\pm y} \equiv \Delta$. Dividing the 
lattice into two sublattices $A$ and $B$ and going to the momentum 
representation, Hamiltonian (\ref{HRVB}) can be diagonalized
\begin{equation}
H_{\text{RVB}} = \sum_{\bbox{k}\nu}
\xi^{\nu}_{\bbox{k}} f^{\dagger}_{\bbox{k}\nu} f_{\bbox{k}\nu},
\label{HRVB2}
\end{equation}
with index $\nu=\pm$. The spectrum of spin excitations or spinons is
\[
\xi^{\pm}_{\bbox{k}} = \pm 2\Delta\left(\cos^2k_x+\cos^2 k_y\right)^{1/2}.
\]
It has nodes at $(\pm\pi/2,\pm\pi/2)$, yielding a 
V-shaped pseudogap in the density of states centered at the spinon chemical 
potential $\mu_s=0$: $\rho^{(0)}(\omega) = |\omega|/D^2$ (defined per spin 
up/down state), where $D=2\sqrt{\pi}\Delta$ is the spinon half-band-width.

To simulate first a nonmagnetic Zn impurity we introduce into Hamiltonian
(\ref{HRVB}) a local chemical potential $\lambda$ acting on site
$\bbox{R}=0$, which by convention lies on sublattice $A$.
In the limit $\lambda\rightarrow\infty$ spinons are 
expelled from that site, creating a vacancy. The Hamiltonian is then
\begin{equation}
H_{\text{Zn}} = H_{\text{RVB}} + \lambda \sum_{\sigma}
f^{\dagger}_{0\sigma} f_{0\sigma}\big|_{\lambda\rightarrow \infty}.
\label{HZN}
\end{equation}
To describe a magnetic Ni impurity, into the empty site we insert
an impurity spin $\bbox{S}_0$ with $S=1$
which is coupled antiferromagnetically to the surrounding Cu spins 
$\bbox{s}_{\bbox{\delta}}$. The corresponding Hamiltonian is 
\begin{equation}
H_{\text{Ni}} =  H_{\text{RVB}} + \lambda \sum_{\sigma}
f^{\dagger}_{0\sigma} f_{0\sigma}\big|_{\lambda\rightarrow \infty} 
+ H_{\text{imp}}
\label{HNI}
\end{equation} 
with the exchange interaction term
\[
H_{\text{imp}} = J'\sum_{\bbox{\delta}} \bbox{S}_0 
\bbox{s}_{\bbox{\delta}}.
\]
Formally, Hamiltonian (\ref{HNI}) differs from Eq.\ (\ref{HZN}) only in 
the presence of an additional term $H_{\text{imp}} \propto J'$. In the 
following, we put emphasis on the case of a magnetic impurity with
$J'>0$. A nonmagnetic impurity can be simulated by setting $J'=0$,
which decouples the impurity site from the rest of the system. 
The $S=1$ impurity spin is then free and can easily be disregarded.
We discuss this limit in the following, but only shortly. More detailed 
treatments on nonmagnetic impurities are given in Ref.\ \onlinecite{KKK97} 
as well as in Refs.\ \onlinecite{NAG95}-\onlinecite{PEP98}.

\section{Local Magnetic Moments}
\label{SEC:LOC}

We analyze an impurity spin $S=1$ embedded in a spin gap 
system as described by Hamiltonian (\ref{HNI}). Spinons stemming from 
the initial Cu spin at site $\bbox{R}=0$ are ejected
by the local potential $\lambda$. The impurity spin, which is
placed in the vacant site, is conveniently represented by two 
spins $\case{1}{2}$, i.e., 
$\bbox{S} = \bbox{S}_a + \bbox{S}_b$. An infinitely strong ferromagnetic 
interaction $H_{c} = -J_c\bbox{S}_a\bbox{S}_b$ between these two spins 
is assumed. Expressing $\bbox{S}_a$ and $\bbox{S}_b$ in terms of
fermionic operators $a_{\sigma}$ and $b_{\sigma}$, respectively,
a mean-field decoupling can be performed:
\begin{equation}
H_{\text{imp}} = -\sum_{\bbox{\delta} \sigma}
\left( \Delta'_{\bbox{\delta}}
\frac{a^{\dagger}_{\sigma} + b^{\dagger}_{\sigma}}{\sqrt{2}}
f_{\bbox{\delta} \sigma} + \text{H.c.} \right)
 - J_c \bbox{S}_{a} \bbox{S}_{b}.
\label{HIM1}
\end{equation}
Introducing operators
$f_{0\sigma} = \left(a_{\sigma} + b_{\sigma}\right)/\sqrt{2}$ and
$d_{\sigma} = \left(a_{\sigma} - b_{\sigma}\right)/\sqrt{2}$, one obtains
\begin{equation}
H_{\text{imp}} = -\sum_{\bbox{\delta} \sigma}
\left( \Delta'_{\bbox{\delta}} f^{\dagger}_{0\sigma}
f_{\bbox{\delta}\sigma} + \text{H.c.} \right)
 - J_{c} \bbox{S}_{\text{eff}} \bbox{s}_0 .
\label{HIM2}
\end{equation}
The impurity spin has thus been decomposed into two $S=\case{1}{2}$
effective spins $\bbox{S}_{\text{eff}}$ and $\bbox{s}_0$. The former
is represented by operators $d_{\sigma}$, the latter by
operators $f_{0\sigma}$. Due to the first term in Eq.\ (\ref{HIM2}), the 
$f$ spinons on the impurity site hybridize with the ones on adjacent 
Cu sites. This process is controlled by the local mean-field parameter 
$\Delta'_{\bbox{\delta}} = J'\sum_{\sigma}\langle f_{\bbox{\delta}\sigma}^
{\dagger} f_{0\sigma}\rangle$ replacing $\Delta_{\bbox{\delta}}$ on 
bonds connecting to the impurity. A system of itinerant spinons 
extending over the whole lattice including the impurity site is formed. 
These itinerant spinons couple ferromagnetically to the localized 
spin $\bbox{S}_{\text{eff}}$. In the presence of a magnetic field 
this coupling is responsible for a polarization of the spinon system
to be discussed in Sec.\ \ref{SEC:SPI}. The $T$ matrix that describes 
scattering of spinons on the localized spin vanishes as $T(\omega) \propto 
\omega\ln|\omega|$ in the flux phase. \cite{KKK97} This means that the 
effective local spin $\bbox{S}_{\text{eff}}$ becomes asymptotically free 
in the limit of low energies. In the remainder of the present section we
analyze this low-energy fixed point, emphasizing the role of 
bond parameters $\Delta'_{\bbox{\delta}}$ that induce an 
inhomogeneity in the spinon sector.

\begin{figure}
\noindent
\centering
\epsfxsize=0.4\linewidth
\epsffile{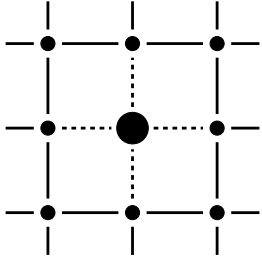}\\[6pt]
\caption{Mean-field parameters $\Delta'_{\bbox{\delta}}$ and 
$\Delta_{\bbox{\delta}}$ are assigned to bonds that do (dashed line) 
or do not (solid line) connect to the impurity site denoted
by a large dot.}
\label{FIG:BON}
\end{figure}

First we consider the special case of equal exchange integrals
$J'=J$. Regarding the spinon sector, the impurity site becomes
indistinguishable from the rest of the system as $\Delta'_{\bbox{\delta}} 
= \Delta_{\bbox{\delta}}$. The spin $\bbox{s}_0$ takes the 
role of the original Cu spin at $\bbox{R}=0$, and a homogeneous spin 
liquid, as described by $H_{\text{RVB}}$, Eq.\ (\ref{HRVB}), is formed. 
Generally, the two exchange integrals differ, $J' < J$, and 
translational invariance of the spinon system is broken. The bond 
parameter then acquires an additional spatial dependence which
has to be treated self-consistently. To simplify the discussion, 
however, we distinguish only between bonds that do and do not 
connect to the impurity (see Fig.\ \ref{FIG:BON}), respectively:
\[
\Delta_{ij} = \left\{ 
\begin{array}{l}
\Delta'_{\bbox{\delta}}\quad\text{for}\quad i=0 \; \text{or} \; j=0\\
\Delta_{\bbox{\delta}}\quad\text{for}\quad i,j\ne 0,
\end{array}\right.
\]
where $\Delta_{\bbox{\delta}}$ is the mean-field parameter of the 
impurity-free system. 
The two parameters
$\Delta'_{\bbox{\delta}}$ and $\Delta_{\bbox{\delta}}$ are assumed to 
exhibit the same phase relation, but in general their amplitudes differ.
As a result, spinons scatter on the impurity bonds. To study this effect
we write the spinon part of Hamiltonian (\ref{HNI}) as
\begin{equation}
H^{\text{sp}}_{\text{Ni}} = H_{\text{RVB}} + 
(1-x) \sum_{\bbox{\delta} \sigma} \left(\Delta_{\bbox{\delta}} 
f^{\dagger}_{0 \sigma} f_{\bbox{\delta} \sigma} + \text{H.c.}
\right),
\label{HSC}
\end{equation}
where $H_{\text{RVB}}$ represents the impurity-free system.
The scattering amplitude $(1-x)$ with $x=|\Delta'_{\bbox{\delta}}/
\Delta_{\bbox{\delta}}|$ is controlled by the ratio of $J'$ to $J$. It 
vanishes for $J'=J$, and has to be treated self-consistently for $J'<J$.
Approximately, we find $x=J'/J$.
At this point, we introduce spinon propagators $g^{(0)}_{\lambda}(i\omega) = 
-\langle T_{\tau} f_{\lambda}(\tau) f^{\dagger}_{\lambda}(0)
\rangle^0_{i\omega} = (i\omega-\xi_{\lambda})^{-1}$ and 
$g_{\lambda\lambda'}(i\omega) = -\langle T_{\tau} f_{\lambda}(\tau) 
f^{\dagger}_{\lambda'}(0)\rangle_{i\omega}$ for the 
pure and impurity-doped system. These can be related by a scattering 
matrix $T_{\lambda \lambda'}(i \omega)$:
\begin{equation}
g_{\lambda\lambda'}(i \omega) = g^{(0)}_{\lambda}(i \omega)
\delta_{\lambda \lambda'} + g^{(0)}_{\lambda}(i \omega) T_{\lambda \lambda'}
(i\omega)g^{(0)}_{\lambda'}(i \omega).
\label{PRO}
\end{equation}
A simplified notation $\lambda = (\bbox{k},\nu)$
and Matsubara frequencies $i\omega = i(2n+1)\pi T$, where $T$ denotes 
temperature and $n$ integer numbers, are employed. The
$T_{\lambda\lambda'}$ matrix in Eq.\ (\ref{PRO}) describes scattering of
spinons on the four bonds that connect the impurity site to its
nearest neighbors. We find it to be given by the expression
\begin{equation}
T_{\lambda \lambda'}(i \omega) = \frac{t_{\lambda \lambda'}
(i\omega)}{i \omega G^{(0)}(i \omega) + p^2},
\label{TMA}
\end{equation}
with
\begin{eqnarray*}
t_{\lambda \lambda'}(i \omega)&=& \frac{1-x}{1+x} G^{(0)}(i \omega)(i
\omega - \xi_{\lambda})(i \omega - \xi_{\lambda'}) \nonumber\\
&&+\frac{x}{1+x}(2i \omega - \xi_{\lambda} - \xi_{\lambda'}) - i\omega.
\end{eqnarray*}
Here $G^{(0)}(i\omega) = \sum_{\lambda} g^{(0)}_{\lambda}(i\omega)
= -(2i\omega/D^2) \ln(D/|\omega|)$, and $p^2 = x^2/(1-x^2)$. The important 
point is that in the flux phase the scattering matrix of Eq.\ (\ref{TMA}) 
has two poles that are determined by the roots of 
\begin{equation}
\omega G^{(0)}(i \omega \rightarrow \omega + i0^+) + p^2 = 0.
\label{BOU}
\end{equation}
One of the poles lies below the spinon chemical potential which signals 
the formation of a spinon bound state. This can be interpreted as follows:
Due to impurity substitution, one Cu spin loses its RVB singlet partner.
In a spin gap system in which short-range spin-singlet correlations dominate, 
this unpaired spin does not dissolve into the RVB ground state but rather
forms a local moment distributed over Cu sites in the proximity of the
impurity. At finite coupling $J'$ this moment forms a local singlet with 
the impurity-site spinon $f_{0\sigma}$. The characteristic binding 
energy $\omega_K$ and lifetime $\delta_K$ of the resulting
bound state are given by the real and imaginary part of the pole,
respectively. For $J'\ll J$, one obtains
\begin{equation}
\omega_K = \frac{\pi}{4} \frac{J'}{\ln D/J'}, \quad
\delta_K = \frac{\pi}{4} \frac{\omega_K}{\ln D/\omega_K}.
\label{WDK}
\end{equation}
In the following, two different energy scales are distinguished: 
$\omega < \omega_K$ and $\omega>\omega_K$. These control the physics 
at large and short distance from the impurity as compared to 
$R_K = D/\omega_K$, respectively, where $R_K$ is measured in
units of lattice spacing. 

\begin{figure}
\noindent
\centering
\epsfxsize=0.42\linewidth
\epsffile{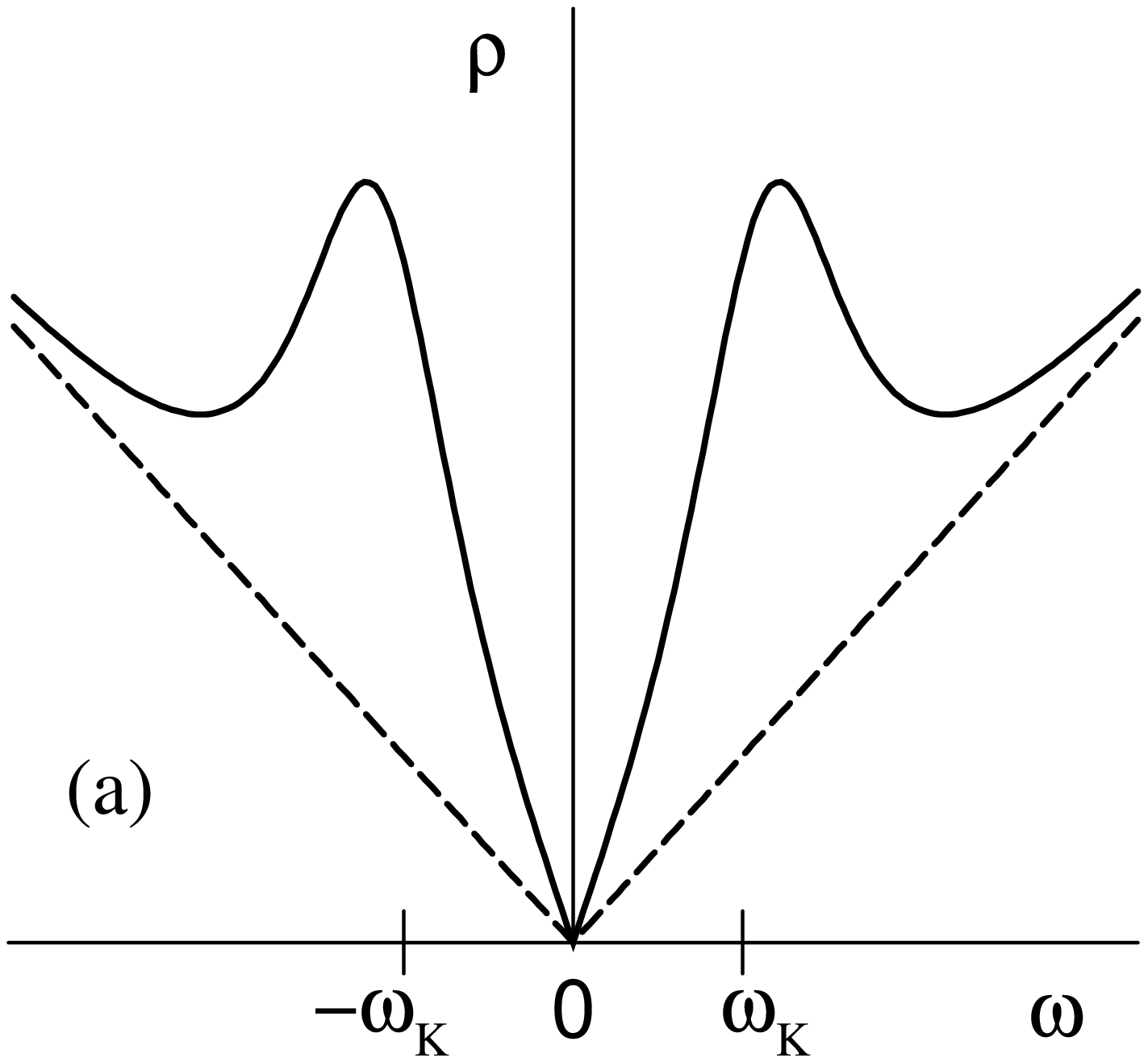}
\hspace{0.05\linewidth}
\epsfxsize=0.42\linewidth
\epsffile{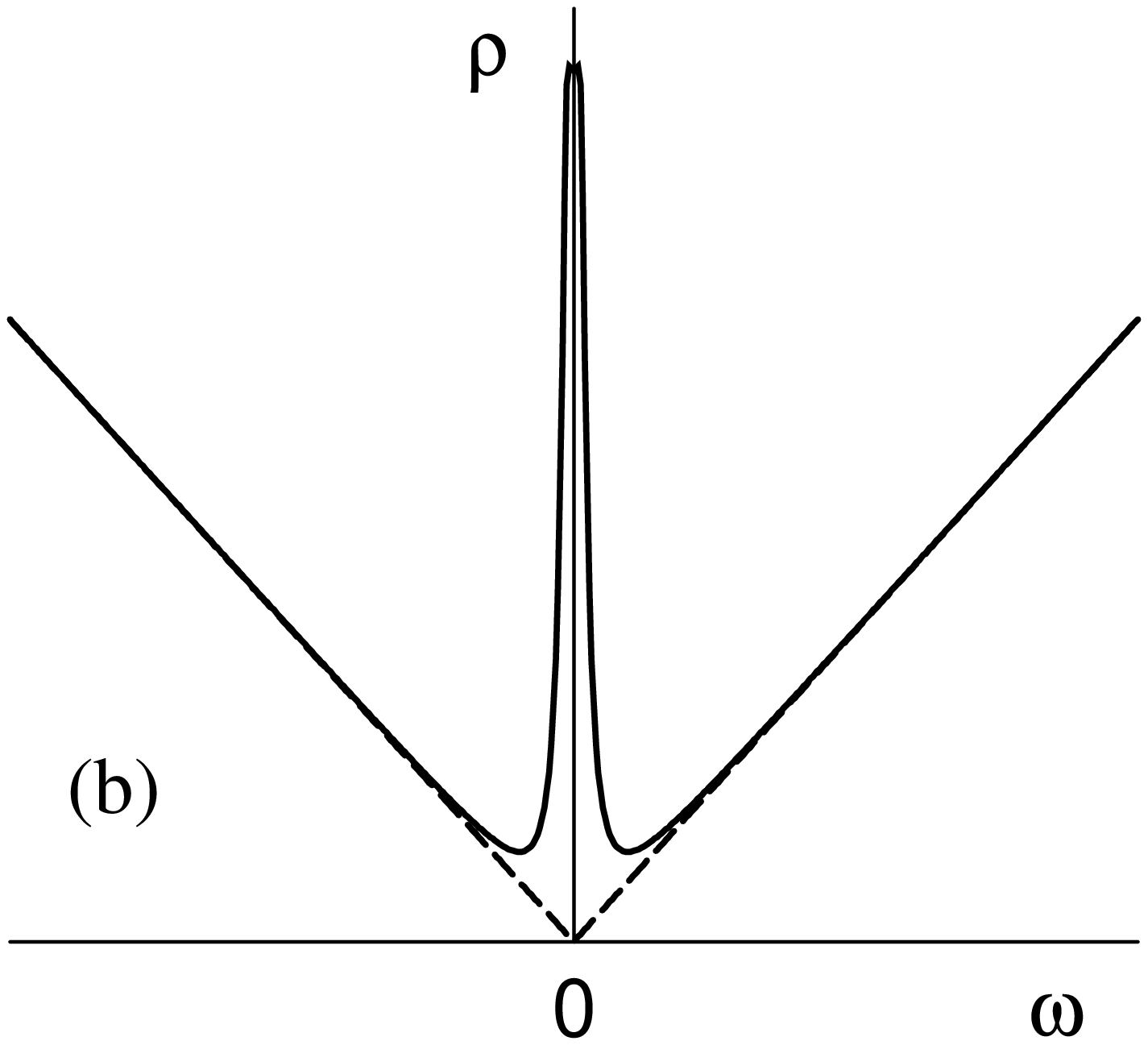}\\[6pt]
\caption{Schematic plot of the spinon density of states for (a) $J'>0$
and (b) $J'=0$ corresponding to magnetic and nonmagnetic impurities,
respectively. Solid lines represent the impurity-doped system,
dashed lines the pure system. The spinon chemical potential $\mu_s$ lies
in the center of the gap, and the $\delta$ function in (b) is artificially 
broadened.}
\label{FIG:DOS}
\end{figure}

First we analyze the low-energy fixed point of the system with a 
magnetic impurity for which $J'$ and hence $\omega_K$ are finite. 
It is determined by the regime $\omega<\omega_K$ and applies to 
distances $R>R_K$ from the impurity site. We calculate 
the impurity contribution $\delta\rho(\omega)$ to the density of 
states from the Green's function $\delta G(i\omega) = 
\sum_{\lambda\lambda'} g_{\lambda}^{(0)}(i\omega) T_{\lambda\lambda'}
(i\omega)g_{\lambda'}^{(0)}(i\omega) = (\partial/\partial i\omega)\ln
[i\omega G^{(0)}(i\omega)+p^2]$. For $\omega\ll D$, the latter is
\begin{equation}
\delta G(i\omega) = \frac{2 G^{(0)}(i\omega)}{i\omega G^{(0)}(i\omega)+p^2},
\end{equation}
which yields
\begin{equation}
\delta \rho (\omega) = \frac{2}{\pi}\omega_K\delta_K
\frac{|\omega|}{(\omega^2-\omega^2_K)^2+(2\omega_K\delta_K)^2}.
\label{DOS1}
\end{equation}
Figure \ref{FIG:DOS}(a) schematically shows the spinon density
of states $\rho^{(0)}(\omega)$ and $\rho(\omega)=\rho^{(0)}(\omega)
+\delta\rho(\omega)$ for the pure and impurity-doped system.
The very existence of a magnetic pseudogap
is found to be unaffected by the presence of the impurity~--
$\rho^{(0)}(\omega)$ as well as $\rho(\omega)$ vanish linearly in 
the limit $\omega\rightarrow 0$. As a consequence, the static spin
susceptibility, which is related to the spinon density of states 
by \cite{REM1}
\begin{equation}
\chi(T) = \frac{1}{4T} \int_{-\infty}^{\infty} dx
\frac{\rho(x)}{\cosh^2(x/2T)},
\end{equation}
vanishes as $\propto T$ at low temperatures. This indicates that in the 
low-energy limit all spins (except $\bbox{S}_{\text{eff}}$ which is not part of 
the spinon system) participate in the formation of singlets. The spinon bound 
state discussed above hence partially screens the impurity spin 
by forming a Kondo singlet with $\bbox{s}_0$. An effective $S=\case{1}{2}$ 
impurity spin $\bbox{S}_{\text{eff}}$ remains. In this 
underscreened Kondo problem, the spinon binding energy $\omega_K$ of 
Eq.\ (\ref{WDK}) plays the role of the Kondo temperature:
$T_K=\omega_K$. 
For temperatures $T\gg T_K$, the susceptibility associated with
the spinon bound state is that of a free spin $\case{1}{2}$,
i.e., $\chi(T)=1/(4T)$; simultaniously, the original $S=1$ impurity spin 
is recovered. We note that the Kondo temperature exhibits 
an unconventional power-law dependence on the coupling parameter $J'$, 
contrasting the conventional exponential behavior. This peculiarity 
is ascribed to the fact that the impurity spin couples to bound 
spinons which are predomenantly in localized rather than bandlike 
states. Finally, we shortly discuss how the presence of 
a Kondo singlet affects the properties of the spinon system at $T\ll T_K$. 
Although the impurity does not fill the magnetic pseudogap, it
nevertheless renormalizes its slope. The leading term in a 
low-energy expansion of Eq.\ (\ref{DOS1}) is related to the density of 
states of the pure system by
\begin{equation}
\delta\rho(\omega) = \frac{1}{p^2} \rho^{(0)}(\omega),
\label{DOS2}
\end{equation}
valid for $J'\ll J$. At low energy and large distance from the impurity, 
the spinon system hence behaves qualitatively as in the impurity-free case.

To finish the discussion of magnetic moments, we turn to the case of a 
nonmagnetic impurity. The relevant physics is modelled by decoupling
the spinon sector from the impurity site, setting $J'=0$, and
by discarding contributions stemming from the impurity spin which is
now free. Since $\omega_K$ consequently vanishes, one is 
always in the regime $\omega>\omega_K$. A Kondo singlet cannot form even 
in the zero-energy limit as the impurity carries no inherent spin. The 
spinon bound state induced by the impurity lies at the spinon 
chemical potential in the center of the pseudogap [see Fig.\ 
\ref{FIG:DOS}(b)]:
\begin{equation}
\delta\rho(\omega) = \delta(\omega).
\end{equation}
This is associated with the magnetic susceptibility $1/(4T)$ of a free 
spin $\case{1}{2}$ which holds down to zero temperature. We note that 
the impurity-induced moment is broadly distributed over planar Cu sites
on sublattice $B$ that does not contain the impurity site, its density falling 
off as $R^{-2}$ with distance from the impurity. \cite{KKK97} 

To summarize, magnetic Ni and nonmagnetic Zn impurities are both associated 
with $S=\case{1}{2}$ magnetic moments. These are, however, of very different 
natures (see Fig.\ \ref{FIG:RVB}): 
\begin{figure}
\noindent
\centering
\hfill
\epsfxsize=0.4\linewidth
\epsffile{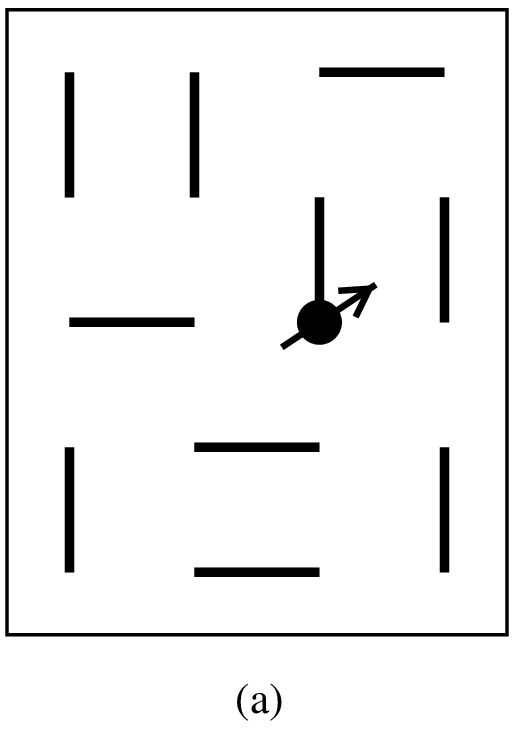}
\hfill
\epsfxsize=0.4\linewidth
\epsffile{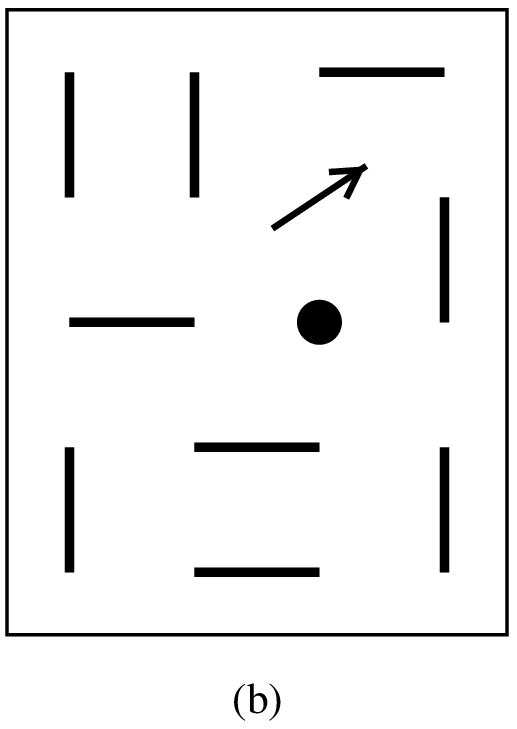}
\hfill\\[6pt]
\caption{``Snapshot'' of the low-energy fixed point of a RVB liquid state 
with (a) the $S=1$ magnetic impurity and (b) the nonmagnetic impurity 
denoted by a dot. 
In the former case, the impurity spin is partially screened by moments 
induced on Cu sites. Effectively, a local impurity spin $\case{1}{2}$ and 
a ``healed'' spin liquid results. In the latter case, the 
impurity induces a broadly distributed moment that resides on Cu sites 
in the proximity of the impurity.}
\label{FIG:RVB}
\end{figure}
In the former case, the spinon bound state partially screens the 
original $S=1$ impurity spin. One is left with an 
effective impurity spin $\case{1}{2}$ ferromagnetically coupled to
an ensemble of inherent spinons that, in the absence of a magnetic
field, behaves qualitatively the same as an impurity-free system. 
In the latter case, the moment is carried by the spinon bound state 
itself, and is broadly distributed over Cu sites.

\section{Spin Polarization}
\label{SEC:SPI}

The effective impurity moments discussed in Sec.\ \ref{SEC:LOC}
can be polarized by 
applying an external magnetic field. In this section we analyze the 
incidental local response of planar Cu spins. In the case of a magnetic 
impurity, the applied field acts on a localized impurity spin $\case{1}{2}$ 
ferromagnetically coupled to the spin liquid. Cu spins respond via a 
RKKY-type interaction. In the case of a nonmagnetic impurity, the moment 
itself resides on Cu sites. Applying a magnetic field therefore directly 
polarizes the Cu spins.

\begin{figure}
\noindent
\centering
\epsfxsize=0.8\linewidth
\epsffile{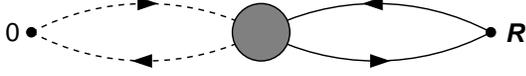}\\[12pt]
\caption{Diagrammatic representation of the static polarization
$K_{\text{Ni}}(T,\bbox{R})$ of a Cu spin at site $\bbox{R}$ due to
RKKY coupling to the localized impurity moment. Dashed
and solid ovals represent particle-hole convolution functions
$\Pi_d(i\varepsilon')$ for the local moment and 
$\Pi_f(i\varepsilon,\bbox{R})$ for itinerant spinons,
respectively. The effective coupling is described by the vertex function 
$J_c(i\varepsilon-i\varepsilon')$ denoted by a circle.}
\label{FIG:BUB}
\end{figure}

We first discuss the situation of a magnetic impurity. The static 
polarizability is defined by $K_{\text{Ni}}(T,\bbox{R}) = \langle 
T_{\tau} s^z_{\bbox{R}}(\tau) S_{\text{eff}}^z(0) \rangle_{\omega=0}$, 
where $s^z_{\bbox{R}}$ and $S_{\text{eff}}^z$ denote the $z$ components 
of a given Cu spin at site $\bbox{R}$ and of the effective impurity spin, 
respectively. It is expressed in terms of Green's functions as (see 
Fig.\ \ref{FIG:BUB})
\begin{equation}
K_{\text{Ni}}(T,\bbox{R}) = -T^2\sum_{\varepsilon,\varepsilon'}
\Pi_d(i\varepsilon') J_c(i\varepsilon-i\varepsilon')
\Pi_f(i\varepsilon,\bbox{R})
\label{POL}
\end{equation}
with particle-hole convolution functions
\begin{eqnarray*}
\Pi_d(i\varepsilon) &=& D^2(i\varepsilon),\\
\Pi_f(i\varepsilon,\bbox{R}) &=& G(i\varepsilon,-\bbox{R}) 
G(i\varepsilon,\bbox{R}).
\end{eqnarray*}
Here the impurity Green's function is $D(i\omega) = -\langle T_{\tau} 
d_{\sigma}(\tau) d^{\dagger}_{\sigma}(0)\rangle_{i\omega} = 1/(i\omega)$, 
and the intersite spinon Green's function $G(i\omega,\bbox{R}) = 
-\langle T_{\tau} f_{0\sigma}(\tau) f^{\dagger}_
{\bbox{R}\sigma}(0)\rangle_{i\omega}$. 
Operators $d$ and $f$ act on separated sectors of the Hilbert space.
At site $\bbox{R}=0$, however, $f$ spinons are polarized by the local spin 
$\bbox{S}_{\text{eff}}$ due to the ferromagnetic interaction of bare
strength $J_c$. This coupling is accounted for by the vertex function 
$J_c(i\omega)$. Employing a ladder approximation it is
\begin{equation}
J_c(i\omega) = \frac{J_c}{1+J_c \Pi_c^{i\omega}} = \frac{1}
{\Pi_c^{i\omega}},
\label{LAD}
\end{equation}
with 
\[
\Pi_c^{i\omega} = -T\sum_{\varepsilon} D(i\varepsilon+i\omega) 
G(i\varepsilon,\bbox{R}=0).
\]
The second equality in Eq.\ (\ref{LAD}) holds due to
$J_c$ being infinitely large. Replacing the vertex function 
by its zero-frequency limit, $J(i\omega)\rightarrow J(0)$,  
the polarizability of Eq.\ (\ref{POL}) can be factorized. Within this 
approximation, which is valid at low temperatures, one obtains
\begin{equation}
K_{\text{Ni}}(T,\bbox{R}) = \chi_{\text{eff}}(T) J_c(0) 
\chi_{\text{pl}}(T,\bbox{R}).
\label{POL2}
\end{equation}
The polarizability has thus been decomposed into the magnetic 
susceptibility of the effectively free $\case{1}{2}$ impurity spin,
$\chi_{\text{eff}}(T)=1/(4T)$, the nonlocal magnetic susceptibility of 
CuO$_2$ planes, $\chi_{\text{pl}}(T,\bbox{R})$, and an effective
coupling parameter $J_c(0)$. The susceptibilities are defined as
$\chi_{\text{eff}}(T) = \langle T_{\tau} S^z_{\text{eff}}(\tau)
S^z_{\text{eff}}(0) \rangle_{\omega=0} =  -T\sum_{\varepsilon} 
\Pi_d(i\varepsilon)$ and $\chi_{\text{pl}}(T,\bbox{R}) = 
\langle T_{\tau} s^z_{\bbox{R}}(\tau) s^z_0(0)\rangle_{\omega=0} = 
-T\sum_{\varepsilon} \Pi_f(i\varepsilon,\bbox{R})$.

To further analyze the polarizability in
Eq.\ (\ref{POL2}), $J_c(0)$ and $\chi_{\text{pl}}(\bbox{R})$ 
have to be evaluated. This requires the on-site and intersite spinon 
Green's functions
\begin{equation}
G(i\omega,\bbox{R}) = \left\{
\begin{array}{l}
\displaystyle
\left(\frac{p}{x}\right)^2 \frac{G^{(0)}(i\omega)}{i\omega G^{(0)}(i\omega)+p^2}
\quad\text{for}\quad R=0\\
\displaystyle
\frac{1}{x} G^{(0)}(i\omega,\bbox{R})\quad\text{for}\quad R>R_K,
\end{array}\right.
\label{GRE}
\end{equation}
where $G^{(0)}(i\omega,\bbox{R})$ is defined for the impurity-free system
\begin{equation}
G^{(0)}(i\omega,\bbox{R}) = -\frac{2i|\omega|}{D^2}
\varphi(\bbox{R}) K_1\left(\frac{R |\omega|}{D}\right),
\label{G0R}
\end{equation}
with a modified Bessel function of the second kind, $K_{\nu}(x)$. 
Equation (\ref{G0R}) 
holds for sites on sublattice $B$; contributions from sublattice $A$ 
containing the impurity are found to be negligible. The angular dependence 
is determined by the phase factor
\begin{equation}
\varphi(\bbox{R}) = \frac{1}{2}\left(\tilde{R}^+e^{i\pi R^+/2} +
\tilde{R}^-e^{i\pi R^-/2}\right)
\end{equation}
with $R^{\pm} = R_x\pm R_y$ and $\tilde{R}^{\pm} = (R_x\pm iR_y)/R$.
We are now in the position to calculate the effective coupling
parameter from the zero-frequency limit of Eq.\ (\ref{LAD}),
\begin{equation}
J_c(0) = \left\{
\begin{array}{l}
D \quad \text{for} \quad J'=J\\
2\omega_K \quad \text{for} \quad J'\ll J,
\end{array}\right.
\label{JC0}
\end{equation}
and the nonlocal spin susceptibility of CuO$_2$ planes
in the presence of the impurity,
\begin{equation}
\chi_{\text{pl}}(\bbox{R}) = 
-\frac{3}{4\pi} \frac{1}{Jx^2} \frac{\Phi(\bbox{R})}{R^3},
\label{SPL}
\end{equation}
the latter being valid for $\bbox{R}\in B$ with $R>R_K$. The phase 
factor in Eq.\ (\ref{SPL}) is defined by $\Phi(\bbox{R}) = 
|\varphi(\bbox{R})|^2$. Finally, combining these results, we 
obtain 
\begin{equation}
K_{\text{Ni}}(T,\bbox{R}) = -\frac{3}{16\pi} \frac{J_c(0)}{Jx^2}
\frac{\Phi(\bbox{R})}{R^3}\frac{1}{T} ,
\label{PNI}
\end{equation}
which describes the polarizability of a Cu spin at site $\bbox{R}\in B$ 
responding to a magnetic field that acts on the effective impurity spin
$\bbox{S}_{\text{eff}}$; contributions
from sublattice $A$ are found to be small. We note that the
$T^{-1}$ Curie behavior displayed by Eq.\ (\ref{PNI}) stems 
solely from the susceptibility $\chi_{\text{eff}}(T)$ of the effective 
impurity spin. Within the present mean-field treatment, the planar 
susceptibility is independent of temperature: $\chi_{\text{pl}}
(T,\bbox{R}) = \chi_{\text{pl}}(\bbox{R})$.

We now briefly review the result for a nonmagnetic impurity
which was derived in Ref.\ \onlinecite{KKK97}. Here, the Cu spins carry 
the impurity-induced moment, and can therefore be directly polarized by 
the magnetic field. The polarizability is given by the local susceptibility
of the impurity-induced moment, $K_{\text{Zn}}(T,\bbox{R}) =
\delta\chi_{\text{pl}}(T,\bbox{R}) = 
\sum_{\bbox{R}'}\left(\langle T_{\tau} s^z_{\bbox{R}}(\tau)
s^z_{\bbox{R}'}(0) \rangle_{\omega=0} - \langle T_{\tau}
s^z_{\bbox{R}}(\tau) s^z_{\bbox{R}'}(0) \rangle_{\omega=0}^0\right)$,
yielding
\begin{equation}
K_{\text{Zn}}(T,\bbox{R}) = \frac{1}{2\pi} \frac{\Phi(\bbox{R})}{R^2}
\frac{1}{T\ln D/T}.
\label{PZN}
\end{equation}
Equation (\ref{PZN}) is valid for $\bbox{R}\in B$, while contributions
from $\bbox{R}\in A$ are again negligible. The polarizability is found
to decay slowly as $R^{-2}$ with distance from the impurity 
which compares to a $R^{-3}$ behavior in the case of Ni, reflecting 
the delocalized nature of the moment induced by a Zn impurity. Further, 
a logarithmic correction to the Curie-like temperature behavior
is to be marked.

In deriving Eqs.\ (\ref{PNI}) and (\ref{PZN}) for the polarizability of Cu 
spins, we have, up to this point, built upon RVB mean-field theory. This
picture accounts well for the spin liquid features of 
underdoped cuprates including the presence of 
a magnetic pseudogap. Its strength lies on the description of long-range 
properties controlled by low-energy excitations. The mean-field treatment 
does, however, severely underestimate local AF correlations which 
reflect the proximity of a critical instability towards AF spin 
ordering. As a consequence, the above expressions contain no reference to 
the AF correlation length which was suggested to introduce a temperature 
dependence beyond the Curie behavior of free moments. \cite{MOR98}
Furthermore, mean-field theory yields a polarizability of Cu spins on one 
sublattice only, undervaluing the staggered magnetization of spins on the 
opposite sublattice. This is in disaccord with NMR measurements \cite{BOB97} 
that yield no overall shift of the $^{17}$O line, as would be
expected from the polarization of only one sublattice as well as with
numerical studies. \cite{MAR98}

To compensate for these deficiencies of the mean-field treatment,
we simulate the closeness of the spin system towards an
antiferromagnetically ordered state by performing a random-phase 
approximation (RPA) in the magnetic susceptibility. In the momentum 
representation, the susceptibility of planar Cu spins then 
becomes
\begin{equation}
\chi_{\text{pl}}^{\text{RPA}}(T,\bbox{q})=\chi_{\text{pl}}(\bbox{q}) 
S(T,\bbox{q})
\label{RPA}
\end{equation}
with the Stoner enhancement factor
\begin{equation}
S(T,\bbox{q}) = \frac{1}{1+J_{\bbox{q}} 
\chi_{\text{pl}}(T,\bbox{q})},
\label{STO1}
\end{equation}
where $J_{\bbox{q}} = 2J(\cos q_x+\cos q_y)$. We closely follow the 
theory of a nearly AF Fermi liquid,\cite{MON94} which maps Eq.\ (\ref{RPA})
onto a phenomenological expression involving the AF correlation length 
$\xi(T)$. Within this picture, $\chi_{\text{pl}}^{\text{RPA}}(T,\bbox{q})$
is assumed to be controlled solely by the momentum region close to
the AF wave vector $\bbox{Q}=(\pi,\pi)$. However, we do take a  
slightly different point of view in this respect: The momentum
dependence of the bare susceptibility $\chi_{\text{pl}}(\bbox{q})$
in Eq.\ (\ref{RPA}),
which describes the long-range characteristics of spin correlations 
in the presence of a magnetic pseudogap, is explicitly kept. Only
the scaling function $S(T,\bbox{q})$, which controls 
short-range AF correlations, is expanded around $\bbox{Q}=(\pi,\pi)$. 
Identifying
$J\chi_{\text{pl}}(T,\bbox{Q})/[1-4J\chi_{\text{pl}}(T,\bbox{Q})]=\xi^2(T)$
and $1/[J\chi_{\text{pl}}(T,\bbox{Q})]=\alpha$, Eq.\ (\ref{STO1}) can
be written in phenomenological form
\begin{equation}
S(T,\bbox{q})=\frac{\alpha\xi^2(T)}{1+(\bbox{q}-\bbox{Q})^2\xi^2(T)},
\label{STO}
\end{equation}
where $\alpha\approx 1$ on a mean-field level. We note that the explicit 
form of $\xi(T)$ lies beyond the accessibility of a mean-field treatment,
and has to be chosen according to general physical considerations.

Turning back to real space, the nonlocal susceptibility is
\begin{equation}
\chi^{\text{RPA}}_{\text{pl}}(T,\bbox{R}) = 
\sum_{\bbox{R}'\in B} \chi_{\text{pl}}(\bbox{R}') S(T,\bbox{R}-\bbox{R}').
\end{equation}
For distances $R\gg\xi(T)$, it can be approximated by
$\chi_{\text{pl}}^{\text{RPA}}(T,\bbox{R}) =
\chi_{\text{pl}} (\bbox{R})\xi^2(T)/2$
with interpolation formula for the $A$ sublattice
$\chi_{\text{pl}}(\bbox{R}\in A) = -(1/z)\sum_{\bbox{\delta}} 
\chi_{\text{pl}}(\bbox{R}+\bbox{\delta})$. Analogous expressions
are obtained for the local susceptibility $\delta\chi_{\text{pl}}
(T,\bbox{R})$ induced by a nonmagnetic impurity.
Combining these results with Eqs.\ (\ref{PNI}) and (\ref{PZN}) and
performing an angular average over phase factors $\Phi(\bbox{R})$,
one finally arrives at the following 
expressions for the polarizability of Cu spins in the impurity-doped system: 
\begin{eqnarray}
K_{\text{Ni}}(T,\bbox{R}) &=& \cos(\bbox{Q}\bbox{R})
\frac{3}{64\pi} \frac{J_c(0)}{Jx^2}\frac{1}{R^3}
\frac{\xi^2(T)}{T},
\label{PNI2}\\
K_{\text{Zn}}(T,\bbox{R}) &=& -\cos(\bbox{Q}\bbox{R})
\frac{1}{8\pi}\frac{1}{R^2}\frac{\xi^2(T)}{T\ln D/T}.
\label{PZN2}
\end{eqnarray}
These equations now hold for both sublattices, $\bbox{R} \in \{A,B\}$,
the staggered nature of spin correlations being manifested in the 
alternating sign implied by $\cos(\bbox{Q}\bbox{R})$. Further, the 
dependence upon the AF correlation length $\xi(T)$ is now explicitly 
accounted for.

\section{NMR Line Broadening}
\label{SEC:NMR}

\begin{figure}
\noindent
\centering
\epsfxsize=0.55\linewidth
\epsffile{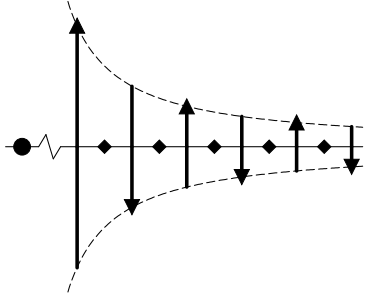}\\[6pt]
\caption{Schematic cut through a CuO$_2$ plane, showing the
position of O ions (diamonds) placed in between successive Cu sites
with staggered spin polarization (arrows). The magnitude of
the polarization falls off with distance from the impurity 
(circle) as $R^{-3}$ in the case of Ni, and as $R^{-2}$ 
in the case of Zn.}
\label{FIG:SPIN}
\end{figure}

The impurity-induced polarization of Cu spins in a magnetic field 
affects the energy levels of nuclear spins via supertransferred hyperfine 
interaction. The coupling of a given nuclear spin $\bbox{I}$ to 
electron spins $\bbox{s}_i$ on close by Cu sites is described by
\begin{equation}
H_{\text{hf}} = \gamma_n\gamma_e C_{\text{hf}}\sum_i\bbox{s}_i\bbox{I},
\end{equation}
where $\gamma_n$ and $\gamma_e$ denote the nuclear and electron gyromagnetic 
ratios, respectively, and $C_{\text{hf}}$ is the supertransferred hyperfine 
coupling constant. In the following, we restrict ourselves to NMR
measurements on $^{17}$O nuclei ($I=\case{5}{2}$). On a mean-field level, 
$\bbox{s}_i$ can be replaced
by its average value $\langle \bbox{s}_i\rangle = K(T,\bbox{R}_i) \bbox{H}_0$
with external magnetic field $\bbox{H}_0$ and polarizability 
$K(T,\bbox{R}_i)$ given by either one of Eqs.\ (\ref{PNI2}) and 
(\ref{PZN2}) for the two types of impurities. Since each $^{17}$O nucleus 
lies symmetrically in between two Cu sites that belong to different 
sublattices with spins polarized in opposite directions, the 
impurity-induced energy shift partially cancels (see Fig.\ \ref{FIG:SPIN}). 
At large enough distance from the impurity, the shift is then
effectively determined by the spatial derivative of the polarizability,
\begin{equation}
\omega(\bbox{R}) = \kappa \frac{\partial |K(T,\bbox{R})|}{\partial R}
\cos\phi,
\label{WRR}
\end{equation}
where $\phi$ denotes the angle enclosed by $\bbox{R}$ and the $x$ 
or $y$ axis
and $\kappa = \gamma_n\gamma_e C_{\text{hf}}H_0$.

\begin{figure}
\noindent
\centering
\setlength{\unitlength}{0.85\linewidth}
\begin{picture}(1.0,0.84)
\put(0,0){
\epsfxsize=\unitlength
\epsffile{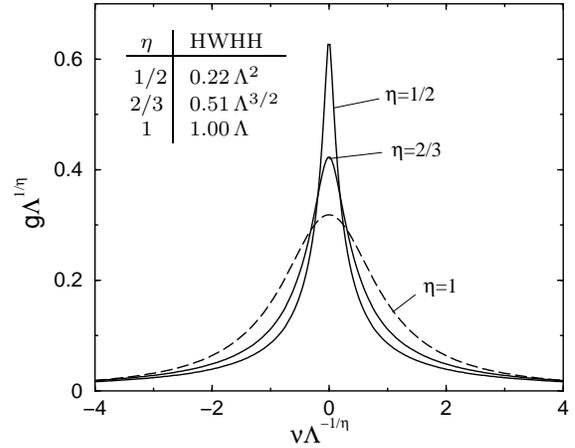}}
\put(0.2,0.75){
\fontsize{0.04\unitlength}{0.045\unitlength}\selectfont
\begin{tabular}[t]{c|@{\protect\rule{0.03\unitlength}{0\unitlength}}l}
$\eta$ & HWHH\\
\hline\rule{0\unitlength}{0.05\unitlength}
$1/2$ & $0.22\,\Lambda^{2}$\\
$2/3$ & $0.51\,\Lambda^{3/2}$ \\
$1$   & $1.00\,\Lambda$
\end{tabular}}
\end{picture}\rule{0.1\linewidth}{0\linewidth}
\caption{Line-shape function $g(\nu)$ obtained by performing a 
Fourier transformation on the characteristic function 
$f(t) = \exp[-\Lambda |t|^{\eta}]$ with $\eta = \case{1}{2}$ for
Ni and $\eta = \case{2}{3}$ for Zn. For comparison, the Lorentzian
line shape of conventional RKKY theory with $\eta=1$ is indicated
by a dashed line. Numerical values for the half width at half height
(HWHH) are given in the inset.}
\label{FIG:POL}
\end{figure}

In a system with randomly distributed impurities of concentration
$c$, the superposition of energy shifts induced
by different impurities leads to a broadening of the NMR line.
We calculate the line shape that follows from Eq.\ (\ref{WRR}), 
employing the formalism of Ref.\ \onlinecite{WAL74}. The line shape 
function $g(\nu)$ is defined as the Fourier transform of the 
characteristic or free-induction function
\begin{equation}
f(t) = \exp\Big[-c\sum_{\bbox{R}} \left(1-e^{i\omega(\bbox{R})t}
\right)\Big].
\end{equation}
Integrating over lattice sites, for Ni and Zn, respectively,
one obtains
\begin{equation}
\ln f(t) = \left\{ \begin{array}{l}
\displaystyle
-\Lambda_{\text{Ni}}|t|^{1/2}\\
\displaystyle
-\Lambda_{\text{Zn}}|t|^{2/3},
\end{array}\right.
\label{LNF}
\end{equation}
with
\begin{eqnarray*}
\Lambda_{\text{Ni}} &=&
\frac{2\sqrt{6}\pi\Gamma(3/4)}{\Gamma(1/4)}\left[\kappa\frac{3}{64\pi} 
\frac{J_c(0)}{Jx^2} \frac{\xi^2(T)}{T}\right]^{1/2}c,\\
\Lambda_{\text{Zn}} &=&
\frac{2\sqrt{3}\pi^2}{\Gamma^2(1/3)} 
\left[\kappa \frac{1}{8\pi} \frac{\xi^2(T)}{T\ln D/T}\right]^{2/3}c.
\end{eqnarray*}
Figure \ref{FIG:POL} shows the different line shapes induced by Ni and Zn 
impurities as obtained by performing a Fourier transformation on 
$f(t)$.  
Comparing to the Lorentzian shape that results from 
$\ln f(t)\propto -|t|$
in conventional RKKY theory, one finds a marked difference in both
shape and width. Using the numerical values shown in the inset
of Fig.\ \ref{FIG:POL}, we finally arrive at the following
expressions for the full linewidth induced by magnetic and
nonmagnetic impurities:
\begin{eqnarray}
\Delta\nu_{\text{Ni}} &=&
2 \times 0.22  \left(\Lambda_{\text{Ni}}\right)^2,
\label{NNI}\\
\Delta\nu_{\text{Zn}} &=&
2 \times 0.51  \left(\Lambda_{\text{Zn}}\right)^{2/3}.
\label{NZN}
\end{eqnarray}

\section{Comparison with Experiment}
\label{SEC:COM}

In this section, we compare the impurity-induced $^{17}$O NMR line 
broadening as described by Eqs.\ (\ref{NNI}) and (\ref{NZN}) with
experimental data of Bobroff {\it et al.} \cite{BOB97} obtained on 
$\text{YBa}_2(\text{Cu}_{1-x}M_x)_3\text{O}_{6.6}$ 
with $M$ = Zn or Ni.

The following constants are chosen: The superexchange 
parameters are specified by $J=0.13$~eV for Cu-Cu interaction and 
$J'=J/2$ for Cu-Ni. \cite{REM2}
A self-consistent treatment yields a scattering 
amplitude $(1-x)=0.5$, where $\Delta=\frac{1}{4}$ has been used.
The Kondo temperature is obtained by numerically solving Eq.\ (\ref{BOU}) 
which gives $T_K = 560$~K. Below this temperature,
the Ni spin is partially screened and behaves as
a spin $\case{1}{2}$ ferromagnetically coupled to the CuO$_2$
plane. The effective coupling constant of this interaction given
by Eq.\ (\ref{JC0}) is $J_c(0) = 0.1$~eV. The hyperfine coupling 
constant between $^{17}$O nuclear and Cu electron 
spins is $C_{\text{hf}}=3.3$~T/$\mu_B$. \cite{MOR98} The 
magnetic-field strength used in the experiment is $H_0=7.5$~T, 
and the concentration of Ni and Zn impurities is $1\%$. The effective 
impurity concentration within CuO$_2$ planes, which is larger by 
a factor of $\case{3}{2}$, is finally $c=1.5\%$.  

Next, an expression for the AF correlation
length $\xi(T)$ has to be specified. It is argued in Ref.\ 
\onlinecite{STO97} that below a critical temperature $T_{\text{cr}}$ 
specified by $\xi(T_{\text{cr}})\approx 2$, the correlation length
assumes the form
\begin{equation}
\xi(T) = \frac{1}{a+bT},
\label{XIT}
\end{equation}
where $a$ and $b$ are fitting constants of the theory. Saturation
of $\xi(T)$ at low temperatures is neglected here. 

\begin{figure}
\noindent
\centering
\epsfxsize=0.85\linewidth
\epsffile{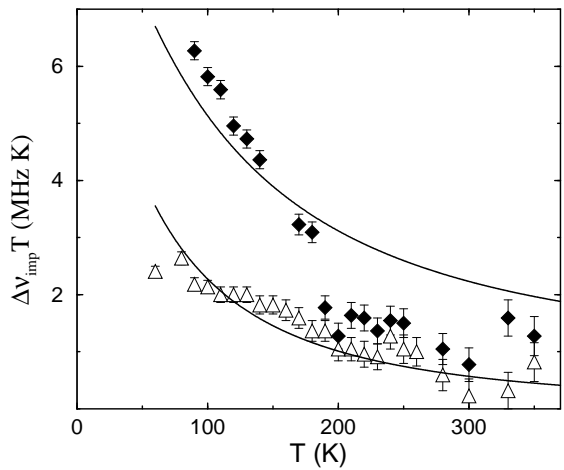}
\caption{Impurity-induced NMR line broadening $\Delta\nu_{\text{imp}}$ 
multiplied by temperature. The
theoretical result is indicated by solid lines fitted to 
experimental $^{17}$O data for $1\%$ Ni-doped (triangles) and
$1\%$ Zn-doped (diamonds) $\text{YBa}_2(\text{Cu}_{1-x}M_x)_3
\text{O}_{6.6}$, $M$ = Ni or Zn.} 
\label{FIG:EXP}
\end{figure}

Figure \ref{FIG:EXP} shows the impurity-induced line broadening $\Delta
\nu_{\text{imp}}$ scaled with temperature.
The curves are fitted to the experimental data by setting 
$a=0.07$ and $b=0.0007$, which correspond to an AF correlation
length of $\xi=4.8$ in units of lattice spacings at $T=200$~K. This
compares well to $\xi=5.9$ obtained in Ref.\ \onlinecite{BAR95}.
No further fitting parameters are needed. 

The theory correctly accounts for the peculiar experimental 
observation of Zn having a more pronounced effect on the NMR
signal than Ni~-- this seems to be in contradiction to SQUID 
measurements on the macroscopic susceptibility. \cite{MEN94} We are able
to ascribe this behavior to the different spatial dependence of
the polarizability: $K(T,\bbox{R})$ decays as $R^{-3}$ in the case of Ni,
but only as $R^{-2}$ in the case of Zn. Averaging over all impurity
site, this leads to an enhanced line-broadening effect of Zn (see 
Fig.\ \ref{FIG:POL}).
Our theory further correctly describes the anomalous non-Curie 
temperature dependence exhibited by the NMR linewidth~-- this
seems to be in disaccord with an almost perfect $T^{-1}$ behavior 
exhibited by the macroscopic susceptibility. \cite{MEN94} One can 
resolve this disagreement by assuming a temperature dependence
of the AF correlation length $\xi(T)$ which enters the polarizability 
of Cu spins. Good agreement with experiment is obtained by employing 
$\xi(T)$ of the form given in Eq.\ (\ref{XIT}).

\section{Conclusion}
\label{SEC:CON}

In summary, we have studied local moments induced in
underdoped cuprates by doping with magnetic ($S=1$) Ni and nonmagnetic
Zn impurities. In the presence of a spin gap,
both types of impurities are associated with $S=\case{1}{2}$ magnetic 
moments in the CuO$_2$ planes. These are, however, of very different 
natures. Ni as well as Zn disturb the spin liquid formed by planar Cu spins, 
resulting in a magnetic moment residing on Cu sites in the proximity of the
impurity. In the case of Ni, this moment partially shields the impurity spin 
below a critical temperature $T_K$ in what 
resembles an underscreened Kondo model; an effective impurity spin 
$\case{1}{2}$ results. Since predominantly localized rather than
bandlike states are involved in the screening of the impurity
spin, the Kondo temperature exhibits an unconventional power-law 
dependence on the coupling constant. 
In the case of Zn, on the other hand, one deals with a $S=\case{1}{2}$ 
moment broadly distributed over Cu sites. 
We have further investigated the RKKY-type response of Cu spins in
a magnetic field. The spin polarization is found to decay as $R^{-3}$ 
with distance from a Ni impurity, but only as $R^{-2}$ in the case of
Zn. This different behavior reflects the delocalized character of 
Zn moments, and explains why Zn has a stronger impact on the NMR
linewidth than Ni. Further, accounting for the presence of 
temperature-dependent AF correlations in underdoped cuprates, we can successfully 
describe the non-Curie behavior of the impurity effect on the NMR linewidth.
In general, it can be concluded that the anomalous impurity properties 
of underdoped cuprates are a clear manifestation of the peculiar mixture of
spin-singlet and antiferromagnetic correlations present in these compounds.



\begin{references}
\bibitem{REM1} Susceptibilities are corrected by a factor of $2$ to
compensate for an underestimation in the present mean-field treatment.
The use of a Majorana fermion representation as in Ref.\ 
\protect\onlinecite{KKK97} makes this correction dispensable.
\bibitem{REM2} The Ni$^{2+}$ ion with spin 1 has two holes on the 
3D shell in $d_{x^2-y^2}$ and $d_{3z^2-r^2}$ orbitals. Since 
superexchange is mainly contributed to by the $d_{x^2-y^2}$ orbital, 
one obtains $J'/J=1/2$.
\end{references}
\end{document}